# Klein-Gordon equation for electrically charged particles with new energy-momentum operator

A.L. Kholmetskii[1,1], T. Yarman[2], O. Missevitch[3]

[1]*Department of Physics, Belarus State University, Minsk, Belarus; alkholmetskii@gmail.com*
[2]*Istanbul Okan University, Istanbul, Turkey & Savronik, Eskisehir, Turkey*
[3]*Institute for Nuclear Problems, Belarus State University, Minsk, Belarus*

**Abstract**

We address the Klein-Gordon equation for a spinless charged particle in the presence of an electromagnetic (EM) field, and focus on its known shortcoming, related to the existence of solutions with a negative probability density. We disclose a principal way to overcome this shortcoming, using our recent results obtained in the analysis of quantum phase effects for charges and dipoles, which prove the need to abandon the customary definition of the momentum operator for a charged particle in an EM field through its canonical momentum, and to adopt the more general definition of this operator through the sum of mechanical and electromagnetic momenta for the system "charged particle in an EM field". We show that the application of the new energy-momentum operator to the Klein-Gordon equation actually eliminates solutions with negative probability density. Some implications of the obtained results are discussed.

## 1. Introduction

Relativistic quantum mechanics includes two fundamental equations: the Klein-Gordon equation, originally conceived for spinless charged particle, as well as the Dirac equation for an electron. In the presence of electromagnetic (EM) field, both equations imply the following definitions of operators of energy and momentum (see, e.g. [1])

$$\hat{E} = i\hbar\frac{\partial}{\partial t} \rightarrow i\hbar\frac{\partial}{\partial t} - e\varphi \; , \; \hat{\boldsymbol{p}} = -i\hbar\nabla \rightarrow -i\hbar\nabla - \frac{e\boldsymbol{A}}{c} \; , \qquad (1a\text{-}b)$$

which ensure the gauge-invariance and the Lorentz invariance. Here, $\hat{E} = i\hbar\frac{\partial}{\partial t}$ is the energy operator, $\hat{\boldsymbol{p}} = -i\hbar\nabla$ is the momentum operator in the absence of EM field, $e$ is the electron charge, $\varphi$, $\boldsymbol{A}$ are the scalar and the vector potentials, correspondingly, and $c$ is the light velocity in vacuum.

In what follows, we mainly focus on the Klein-Gordon equation, which in the presence of EM field reads as [2]

$$\left(i\hbar\frac{\partial}{\partial t} - e\varphi\right)^2 \Psi(\boldsymbol{r},t) = {}^2\left(-i\hbar\nabla - \frac{e\boldsymbol{A}}{c}\right)^2 \Psi(\boldsymbol{r},t) + m^2 c^4 \Psi(\boldsymbol{r},t), \qquad (2)$$

where $\Psi(\boldsymbol{r},t)$ is the wave function of spinless charged particle $e$ with the rest mass $m$.

A serious difficulty related to eq. (2) is the emergence of solutions with a negative probability density $\rho(\boldsymbol{r})$ for a charged particle, which does not have a reasonable physical interpretation. This fact can be easily demonstrated in the case of time-independent EM field, where the probability density does not depend on time, too. For such stationary states, the var-

---

iables $r$ and $t$ are separated from each other, and at a motion of charged particle $e$ in the electric field of the immovable point-like charged source particle, characterized with the scalar potential { and the vector potential $A$=0, we get the solutions (see, e.g. [3])

$$\Psi(\mathbf{r},t)=u(\mathbf{r})e^{-iEt/\hbar},\qquad(3)$$

$$\rho(\mathbf{r})=\left(\frac{E-e\{}{mc^2}\right)\overline{u}(\mathbf{r})u(\mathbf{r}),\qquad(4)$$

where $E$ is the energy of the moving particle, { is the scalar potential produced by the host charge, and $\overline{u}$ denotes the complex conjugate of $u$.

One can see from eq. (4) that at very small distances $r$ between moving charged particle $e$ and an immovable point-like host charge, the probability density $\rho(\mathbf{r})$ for particle $e$, in general, can acquire negative values, which is obviously senseless from the physical viewpoint.

This result indicates that the Klein-Gordon equation for charged particles in an EM field occurs inconsistent. Nevertheless, it was emphasized in [4] that this equation still remains applicable to quantum theory of scalar fields, though some further attempts of its application to charged particles (e.g., with two-component wave function) are also known (see, e.g. [5]).

However, the crucial question – *What is the deep reason, which makes inconsistent the Klein-Gordon equation for a spinless charged particle?* – remains unanswered for many decades.

All the same, as we will show below, the firsts step towards answering this question was made at the beginning of the 21[th] century in the area of physics, which seems to lie far from any problems related to the Klein-Gordon equation: the experimental discovery of two quantum phase effects for dipoles: the Aharonov-Casher (A-C) phase for a moving magnetic dipole [6], and He, McKellar and Wilkens (HMW) phase for a moving electric dipole [7].

We remind that, before the performance of these experiments, an existence of the A-C phase for the moving magnetic dipole in an electric field has been predicted in [8], while the existence of the HMW phase for the moving electric dipole in a magnetic field has been predicted in [9, 10], and both of these predictions were made by using particular expressions for corresponding Lagrangians, describing magnetic/electric dipoles in an EM field. As the result, the actual physical meaning of the A-C and HMW phase effects was not clarified even after their experimental discoveries [6, 7].

In our recent papers [11-15], we suggested three new principal steps:

- to involve an explicitly covariant expression for the Lagrangian of electric/magnetic dipole in an EM field, which allowed us to find a full set of quantum phase effects for moving dipoles, next to the A-C and HMW phases;

- to disclose the physical meaning of quantum phase effects for dipoles as the superposition of the corresponding quantum phases for their point-like charges;

- to show the existence of two new quantum phases for charges, named by us as the complementary electric and complementary magnetic A-B phases, which allow to explain the full set of quantum phase effects for moving dipoles through the introduced principle of Superposition of Quantum Phases (abbreviated below as the SQP). Here, we emphasize that the validity of this principle directly follows from the linearity of fundamental equations of quantum mechanics.

For convenience, in section 2 we shortly reproduce these results and show that the obtained full set of quantum phase effects for charged particles, moving in EM field, cannot be derived from the Schrödinger equation or other fundamental equations of quantum mechanics, when the customary definition of the momentum operator

$$\hat{\boldsymbol{p}} = -i\hbar\nabla \rightarrow \hat{\boldsymbol{p}}_M + \frac{e\hat{\boldsymbol{A}}}{c} \qquad (5)$$

is applied. (Hereinafter $\boldsymbol{p}_M$ denotes the mechanical momentum of charged particle).

Simultaneously we emphasize that the necessary condition for inclusion of the complementary electric and complementary magnetic A-B phases into solutions of fundamental equations of quantum mechanics is the adoption of the new definition of the momentum operator

$$\hat{\boldsymbol{p}} = -i\hbar\nabla \rightarrow \hat{\boldsymbol{p}}_M + \hat{\boldsymbol{p}}_{EM}, \qquad (6)$$

where $\boldsymbol{p}_{EM}$ is the interactional EM momentum for charged particle in an EM field.

Next, in section 2, we explicitly derive the full set of quantum phase effects for a charged particle in an EM field in the weak relativistic limit, using the new momentum operator (6) in the Hamiltonian of the Schrödinger equation.

Considering the strong relativistic case, covered by the Dirac equation and the Klein-Gordon equation, and defining the appropriate modifications of these equations, aimed to include into their solutions the complementary electric and magnetic A-B phases, we have a single option left out, which ensures both the relativistic invariance and the gauge invariance: to replace the four-vector $\{\varphi, \boldsymbol{A}\}$ by the four-vector $\{U_{EM}/c, \boldsymbol{p}_{EM}\}$, where $U_{EM}$ being the interactional EM energy for the system "charged particle in an EM field". Hence, in the presence of EM field, eqs. (1a-b) should be modified to the form

$$i\hbar\frac{\partial}{\partial t} \rightarrow i\hbar\frac{\partial}{\partial t} - U_{EM}, \quad \hat{\boldsymbol{p}} \rightarrow \hat{\boldsymbol{p}} - \hat{\boldsymbol{p}}_{EM}, \qquad (7a\text{-}b)$$

in the Klein-Gordon equation for electrically charged particle, which in the four-form reads as [15]

$$i\hbar\partial^{\nu} \rightarrow i\hbar\partial^{\nu} - \hat{p}^{\nu}_{EM}. \qquad (7c)$$

Here $\nu$=0...3, and $\hat{p}^{\nu}_{EM}$ stands for the operator of EM four-momentum.

The application of new definitions of energy and momentum operators (7a-b) in the Klein-Gordon equation essentially affects its solutions and their physical implications compared to the case of old definitions of these operators (1a-b). In the present paper, we will focus on a particular, but fundamentally important problem of the motion of an electrically charged particle in the electric field of an immovable point-like host charge. We will show that the application of eqs. (7a-b) instead of eqs. (1a-b) does in effect modify the solution of the standard Klein-Gordon equation (4) to a new form, which ensures a positive probability density $\rho(\boldsymbol{r})$ in the entire space. By such a way, we overcome the annoying obstacle to the application of the Klein-Gordon equation to electrically charged scalar particles.

We conclude in section 4.

## 2. Quantum phases for dipoles and point-like charges indicate the necessity to re-define the momentum operator

By the end of the 20[th] century, there were known the electric and magnetic Aharonov-Bohm (A-B) quantum phase effects for point-like charges [16], as well as the A-C [8] and HMW [9, 10] phase effects for magnetic and electric dipole, correspondingly, and all these phases have been observed [17, 6, 7].

At the same time, the application of particular expressions for the Lagrangian of dipoles in the derivation of the A-C and HMW phases left unanswered the question of the possible existence of more quantum phase effects for dipoles and their physical meaning.

This situation motivated us [11, 12] to continue further studies on quantum phase effects for dipoles on the basis of the general covariant expression for the Lagrangian density of a material medium in an EM field [18]

$$L = M^{-\epsilon} F_{-\epsilon} \big/ 2 \,, \tag{8}$$

where $M^{-\epsilon}$ is the magnetization-polarization tensor, and $F_{-\epsilon}$ is the tensor of EM field.

Integrating the Lagrangian density (8) over the volume of a compact dipole, we obtain the Lorentz-invariant Lagrangian [11],

$$L = \boldsymbol{p} \cdot \boldsymbol{E} + \boldsymbol{m} \cdot \boldsymbol{B} \tag{9}$$

with a subsequent derivation of a new motional equation and a new Hamilton function for the dipole [11, 12]. Here $\boldsymbol{p}$ ($\boldsymbol{m}$) stands for the electric (magnetic) dipole moment, and $\boldsymbol{E}$, $\boldsymbol{B}$ designate the electric and magnetic fields, correspondingly. Further transition to the quantum limit allowed obtaining a complete expression for the quantum phase of the dipole [11, 12]. Excluding the Stark phase [19] and the Zeeman phase [20], which do not explicitly depend on the velocity of dipole $\boldsymbol{v}$, we obtain four velocity-dependent components [11, 12], which in the weak relativistic limit, up to the terms $(v/c)^3$, read as:

$$\mathsf{u}_{\text{dipole}}(\boldsymbol{v}) \approx \frac{1}{\hbar c} \int (\boldsymbol{m} \times \boldsymbol{E}) \cdot ds - \frac{1}{\hbar c} \int ((\boldsymbol{p} \times \boldsymbol{B})) \cdot ds - \frac{1}{\hbar c^2} \int (\boldsymbol{p} \cdot \boldsymbol{E}) \boldsymbol{v} \cdot ds - \frac{1}{\hbar c^2} \int (\boldsymbol{m} \cdot \boldsymbol{B}) \boldsymbol{v} \cdot ds \,, \tag{10}$$

where all quantities are evaluated in the laboratory frame.

The identification of four quantum phases for a moving electric/magnetic dipole definitely actualizes the problem of their physical interpretation on the basis of some general and universal approach applicable to each of the phases.

Such a universal approach has been proposed in [13-15], where we emphasized that, due to the linearity of the fundamental quantum mechanical equations, the physical meaning of quantum phases for a moving dipole should be understood through a superposition of corresponding quantum phases for all electric charges composing the dipole.

This novel idea, however, immediately brought up a serious difficulty, related to the fact the magnetic A-B phase – a sole quantum phase previously known for a point-like charge with an explicit dependence on its velocity $\boldsymbol{v}$ – obviously could not explain the origin of four quantum phases for moving dipoles in eq. (10). Therefore, if one adopts the correctness of the SQP principle, then one has to recognize the existence of more, previously unknown, quantum phase effects for point-like charged particles, in addition to the electric and magnetic A-B phases.

In a deeper insight to this problem, it is convenient to express all terms of eq. (10) though the scalar $\{$ and vector $\boldsymbol{A}$ potentials, commonly used in the description of A-B phases for point-like charges. This problem has been solved in [13, 14], where the following expressions were obtained for each term of eq. (10):

$$\mathsf{u}_{mE} = \frac{1}{\hbar c} \int (\boldsymbol{m} \times \boldsymbol{E}) \cdot ds = -\frac{1}{\hbar c^2} \iint_{L\,V} \{ \boldsymbol{j} \cdot ds dV \,, \text{ (A-C phase)} \tag{11a}$$

$$\mathsf{u}_{pB} = -\frac{1}{\hbar c} \int ((\boldsymbol{p} \times \boldsymbol{B})) \cdot ds = -\frac{1}{\hbar c^2} \iint_{L\,V} ... \boldsymbol{A} \cdot ds dV \,, \text{ (HMW phase)} \tag{11b}$$

$$\mathsf{u}_{pE} = -\frac{1}{\hbar c^2} \int (\boldsymbol{p} \cdot \boldsymbol{E}) \boldsymbol{v} \cdot ds = -\frac{1}{\hbar c^2} \iint_{L\,V} ... \{ \boldsymbol{v} \cdot ds dV \,, \tag{11c}$$

$$\mathsf{u}_{mB} = -\frac{1}{\hbar c^2}\int(\boldsymbol{m}\cdot\boldsymbol{B})\boldsymbol{v}\cdot ds = -\frac{1}{\hbar c^3}\iint_{L\,V}(\boldsymbol{j}\cdot\boldsymbol{A})\boldsymbol{v}\cdot ds\, dV\,; \qquad (11d)$$

here ... is the charge density, and $\boldsymbol{j}=...\boldsymbol{u}$, $\boldsymbol{u}$ being the flow velocity of carriers of current of a magnetic dipole defined in its rest frame.

Considering the above eqs. (11a-d), we first address to the HMW phase (11b) and conclude that its representation through the vector potential immediately reveals its physical meaning as a superposition of magnetic A-B phases for each charge, composing the dipole.

This result is important for the validation of the entire SQP principle, given that the existence of the magnetic A-B phase and the existence of the HMW phase have already been proved experimentally (see [17] and [7], correspondingly).

Further comparison of eq. (11a) for the A-C phase with eq. (11c) for the $\mathsf{u}_{pE}$ phase reveals their common origin, where both phases can be presented as a superposition of fundamental phases for point-like charges composing the dipole

$$\mathsf{u}_{c\zeta} = \frac{1}{\hbar c^2}\int e\{\boldsymbol{v}\cdot ds\,, \qquad (12)$$

which we named in [13, 14] as the complementary electric A-B phase.

Next, considering the phase $\mathsf{u}_{mB}$ (11d), and using again the equality $\boldsymbol{j}=...\boldsymbol{u}$, we find that this phase originates from the fundamental phases for each charge of the dipole

$$\mathsf{u}_{cA} = -\frac{e}{\hbar c^3}\int(\boldsymbol{v}\cdot\boldsymbol{A})\boldsymbol{v}\cdot ds\,. \qquad (13)$$

In [13, 14], we suggested to name $\mathsf{u}_{cA}$ as the complementary magnetic A-B phase.

Thus, the application of SQP to the analysis of quantum phase effects for moving electric/magnetic dipoles reveals the existence of two new fundamental quantum phases (12), (13) for point like charges, which, being added to the known electric A-B phase

$$\mathsf{u}_{\zeta} = -\frac{1}{\hbar}\int e\{dt \qquad (14)$$

and magnetic A-B phase,

$$\mathsf{u}_{A} = \frac{1}{\hbar c}\int eA\cdot ds \qquad (15)$$

yield the new expression for the total quantum phase of a charged particle in an EM field

$$\mathsf{u}_{EM} = -\frac{1}{\hbar}\int e\{dt + \frac{e}{\hbar c}\int\boldsymbol{A}\cdot ds + \frac{e}{\hbar c^2}\int\{\boldsymbol{v}\cdot ds - \frac{e}{\hbar c^3}\int\boldsymbol{v}(\boldsymbol{A}\cdot\boldsymbol{v})\cdot ds\,. \qquad (16)$$

Next, we emphasize that our disclosure of complementary electric (12) and complementary magnetic (13) A-B phases requires re-analyzing the entire procedure with respect to an adequate description of quantum phase effects for electrically charged particles, defined by the general expression

$$\mathsf{u} = -\frac{1}{\hbar}\int(\hat{H} - \hat{H}_0)dt\,. \qquad (17)$$

where $\hat{H}$ ($\hat{H}_0$) is the Hamiltonian of a charged particle in the presence (absence) of an EM field.

The necessity for such a re-analysis follows from the known fact that eq. (17) with the standard Hamiltonian of a charged particle in an EM field [21]

$$\hat{H} = \frac{\left(-i\hbar\nabla - e\boldsymbol{A}/c\right)^2}{2m} + e\{ \tag{18}$$

with the standard definition of the momentum operator (5), yields only electric (14) and magnetic (15) A-B phases, and it fails to describe the complementary electric (12) and complementary magnetic (13) A-B phases. (Hereinafter, all variables in the Hamiltonian are understood as operators).

Moreover, one can realize that the Dirac equation and the Klein-Gordon equation also fail to describe the complementary A-B phases (12), (13) with the momentum operator (5).

Therefore, our subsequent problem is to re-define the momentum operator in an appropriate way, where complementary electric (12) and complementary magnetic (13) A-B phases are included into the solutions of the fundamental equations of quantum mechanics, described through the Schrödinger equation for a charged particle in an EM field in a weak relativistic limit, and through the Dirac equation and the Klein-Gordon equation in the relativistic case.

As mentioned above, one of the necessary conditions for covariance of the latter equations is the known fact that the $\{$ and $\boldsymbol{A}$ compose a four-vector. Thus, the appropriate modification of the Hamiltonian for inclusion of the complementary A-B phases (12), (13) into the solution of the fundamental equations of relativistic quantum mechanics leaves only one option: to replace the four-vector $\{\{, \boldsymbol{A}\}$ by the four-vector $\{U_{EM}/c, \boldsymbol{P}_{EM}\}$, which also keeps the gauge invariance of these equations; hence we arrive at eqs. (7a-b).

Further on, let us show, for simplicity in the weak relativistic limit, described by the Schrödinger equation for a charged particle in an EM field, that the redefinition of the momentum operator (6) represents a necessary and sufficient condition for describing the new quantum phases (12) and (13).

First, we have to explicitly derive the corresponding Hamiltonian. Here we take into account that in the classical Hamilton function the interactional EM energy contains only the electric energy component $U_{EM}=e\{$ and does not include the magnetic energy component. This reflects the known result of classical physics that the magnetic component of the Lorentz force does not deliver work. Since the term $U_{EM}=e\{$ is already present in the standard Hamiltonian [21] of the Schrödinger equation, it remains sufficient to replace the vector potential $(e\boldsymbol{A}/c)$ with the interactional EM momentum $\boldsymbol{p}_{EM}$ due to the re-definition of the momentum operator from (5) to (6).

Now let us show that the redefinition of the momentum operator (6) actually represents a sufficient condition to describe new quantum phases (12), (13).

Indeed, with the momentum operator (6), the Hamiltonian of the Schrödinger equation for a charged particle in an EM field takes the form

$$\hat{H} = \frac{\left(-i\hbar\nabla - \boldsymbol{p}_{EM}\right)^2}{2M} + e\{ \ . \tag{19}$$

Presenting the operator $-i\hbar\nabla$ as $m\boldsymbol{v}$ in cross terms of eq. (19), and using the Coulomb gauge, where the operators $\boldsymbol{v}$ and $\boldsymbol{A}$ commute with each other, we obtain with sufficient accuracy $c^{-3}$

$$H = -\frac{\hbar^2}{2m}\Delta - \frac{\boldsymbol{p}_{EM}\cdot\boldsymbol{v}}{c} + e\{ \ , \tag{20}$$

where we have neglected the term $e^2 P_{EM}^2/2Mc^2$, which is warranted in any practical situation.

Further on, taking into account the expression for the Hamiltonian of a particle in the absence of fields (see, e.g. [21]),

$$H = -\frac{\hbar^2}{2m}\Delta , \qquad (21)$$

and combining eqs. (17), (20) and (21), we obtain the total quantum phase of a charged particle in the form

$$\textsf{u}_{EM} = -\frac{1}{\hbar}\int e\{\,dt + \frac{1}{\hbar}\int \boldsymbol{p}_{EM}\cdot d\boldsymbol{s} . \qquad (22)$$

In order to provide a standard representation of quantum phase (23) in terms of scalar and vector potentials, one has to define the interactional EM momentum $\boldsymbol{p}_{EM}$ as a function of $\{$ and $\boldsymbol{A}$.

The solution of this problem is given in [14] with correction in [22]:

$$\boldsymbol{p}_{EM} = \frac{1}{4fc}\int_V (\boldsymbol{E}\times\boldsymbol{B}_e)dV + \frac{1}{4fc}\int_V (\boldsymbol{E}_e\times\boldsymbol{B})dV = \frac{e\boldsymbol{A}}{c} + \frac{ve\{}{c^2} - \frac{ev(\boldsymbol{A}\cdot\boldsymbol{v})}{c^3}. \qquad (23)$$

Here $\boldsymbol{E}_e$, $\boldsymbol{B}_e$ denote the electric and magnetic fields of a charged particle, $\boldsymbol{E}$, $\boldsymbol{B}$ stand for the external electric and magnetic fields, correspondingly, and $V$ designates the entire free space.

Further, substituting eq. (23) into eq. (22), we arrive at eq. (16) for the total quantum phase of a charged particle in an EM field, obtained above through application of SQP to moving dipoles and charges.

By such a way, we have shown that the re-definition of the momentum operator in the form (6) does allow including complementary electric (12) and complementary magnetic (13) A-B phases into the solutions of the Schrödinger equation.

Thus, the results presented in this section definitely indicate the need to abandon the customary definition of the momentum operator (5) towards its new definition (6), which thus should be applied not only to the Schrödinger equation for a charged particle in an EM field, but to fundamental equations of relativistic quantum mechanics, too, to ensure the unification of the basic approaches to the fundamental equations of quantum physics.

The application of the momentum operator (6) to the Dirac equation was shortly considered in [15], both for a charged particle freely moving in an EM field and for electrically bound charges.

In the next section, we will focus on the Klein-Gordon equation with redefined energy and momentum operators (7a-b).

## 3. Klein-Gordon equation with redefined energy and momentum operators for charged scalar particles in EM field

We start with the case of a freely moving spinless particle, where the Klein-Gordon equation reads as (see, e.g., [2, 3, 23])

$$-\hbar^2\frac{\partial^2 \textit{Œ}(\boldsymbol{r},t)}{\partial t^2} = -\ ^2\hbar^2\nabla^2 \textit{Œ}(\boldsymbol{r},t) + m^2c^4 \textit{Œ}(\boldsymbol{r},t), \qquad (24)$$

which results from the relativistic relationship $E^2 - c^2p^2 = m^2c^4$ under the replacements $E \to i\hbar\partial/\partial t$, $\boldsymbol{p} \to -i\hbar\nabla$. Here, the wave function $\textit{Œ}(\boldsymbol{r},t)$ describes a scalar particle at the equality $\textit{Œ}(\boldsymbol{r},t) = \textit{Œ}(-\boldsymbol{r},t)$, and a pseudo-scalar particle at $\textit{Œ}(\boldsymbol{r},t) = -\textit{Œ}(-\boldsymbol{r},t)$.

The probability density defined by eq. (24) is given by the relationship

$$\rho(\boldsymbol{r}) = \frac{E}{mc^2}\Psi^*\Psi \ , \tag{25}$$

which, in general, can be either positive, or negative, since $\Psi$ and $\partial\Psi/\partial t$ can be set arbitrarily and independently on each other due to the presence in eq. (24) of the second time derivative. However, it is obvious that the negative probability density (25) is meaningless from a physical viewpoint. Nevertheless, it was understood (see, e.g., [3]) that for a freely moving particle in the absence of EM fields, the possibility to interpret the function $\rho(\boldsymbol{r})$ as a probability density can still make sense due to the fact that under the choice of a positive solution for the energy

$$E = +\sqrt{\ ^2 p^2 + m^2 c^4} \ , \tag{26}$$

it always remains positive in the absence of external perturbations.

Next, one can see that at positive energy (26), the probability density (25) is also positive, and remains positive forever. Further details for description of a freely moving particle through the Klein-Gorton equation can be found, e.g., in [3].

The actual difficulties in the physical interpretation of the Klein-Gordon equation emerge, when we consider the motion of a spinless charged particle in the presence of an EM field. In this case, for the standard energy and momentum operators (1a-b), eq. (24) is modified to the form (2), which, in the stationary case (which is only considered here for simplicity) yields solutions (3) and (4) for the wave function and probability density, correspondingly.

One can see from eq. (4) that at very small distances $r$ between a charged particle and a point-like source of scalar potential, the probability density $\rho(\boldsymbol{r})$, in general, can acquire negative values, which is obviously senseless from the physical viewpoint.

Just this result was considered for many years as a principle obstacle to the application of the Klein-Gordon equation to the description of electrically charged scalar particles.

In what follows, we would like to demonstrate that such a physically meaningless result is eliminated when we abandon the standard definitions of the energy and momentum operators (1a-b) in favor of their new definitions (7a-b), substantiated above.

In this case, the Klein-Gordon equation acquires the form,

$$\left(i\hbar\frac{\partial}{\partial t} - U_{EM}\right)^2 \Psi(\boldsymbol{r},t) = \ ^2\left(-i\hbar\nabla - \boldsymbol{p}_{EM}\right)^2\Psi(\boldsymbol{r},t) + m^2 c^4\Psi(\boldsymbol{r},t), \tag{27}$$

and here we focus on the particular case where a charged particle is moving in an EM field, created by a source point-like particle, resting at the origin of coordinates. This problem has already been mentioned in the introduction section, and its solution with respect to the standard Klein-Gordon equation (2) is given by eqs. (3), (4). We also noted that eqs. (3), (4), in general, admit a negative probability density $\rho$ of the moving charged particle in the vicinity of the source of EM field, which is meaningless.

Now we solve eq. (27), using explicit expressions for the interactional EM energy and momentum for the considered problem:

$$U_{EM} = e\varphi \ , \ \ \boldsymbol{p}_{EM} = \frac{\boldsymbol{v}e\varphi}{c^2} \ , \tag{28a-b}$$

where in the derivation of the interactional EM momentum (28b) we have used eq. (23) at $\boldsymbol{A}=0$.

Substitution of eqs. (28a-b) into eq. (27) yields

$$\left(i\hbar\frac{\partial}{\partial t}-e\varphi\right)^2\Psi\left(\boldsymbol{r},t\right)= {}^2\left(-i\hbar\nabla-\frac{\boldsymbol{v}e\varphi}{c^2}\right)^2\Psi\left(\boldsymbol{r},t\right)+m^2c^4\Psi\left(\boldsymbol{r},t\right). \tag{29}$$

By squaring both expressions in brackets and using the expressions for energy $E=\gamma mc^2$ and momentum $\boldsymbol{p}=\gamma m\boldsymbol{v}$ in cross-terms (where $\gamma$ is the Lorentz factor, and $\boldsymbol{v}$ is the velocity of the moving particle), we re-arrange eq. (29) as

$$\left(i\hbar\frac{\partial}{\partial t}-\frac{e\varphi}{\gamma}\right)^2\Psi\left(\boldsymbol{r},t\right)= {}^2\left(-i\hbar\nabla\right)^2\Psi\left(\boldsymbol{r},t\right)+m^2c^4\Psi\left(\boldsymbol{r},t\right). \tag{30}$$

One can see that in the considered case $\boldsymbol{A}=0$, eq. (30) differs from eq. (2) by the replacement

$$\varphi \rightarrow \varphi/\gamma. \tag{31}$$

Therefore, the same replacement (31) should be made in its solution compared to eqs. (3) and (4), *i.e.*,

$$\Psi\left(\boldsymbol{r},t\right)=u\left(\boldsymbol{r}\right)e^{-iEt/\hbar}, \tag{32}$$

$$\rho\left(\boldsymbol{r}\right)=\frac{E-\dfrac{e\varphi}{\gamma}}{mc^2}\overline{u}\left(\boldsymbol{r}\right)u\left(\boldsymbol{r}\right). \tag{33}$$

Classically, the Lorentz factor for a charged particle moving in the presence of a scalar potential $\varphi$ is defined by the equality

$$\gamma =\frac{mc^2+e\varphi}{mc^2}. \tag{34}$$

Substituting eq. (34) into eq. (31), we obtain the probability density in the form

$$\rho\left(\boldsymbol{r}\right)=\left(\frac{E}{mc^2}-\frac{e\varphi}{e\varphi+mc^2}\right)\overline{u}\left(\boldsymbol{r}\right)u\left(\boldsymbol{r}\right). \tag{35}$$

One can see that the probability density in eq. (35) is always positive, since $E/mc^2\geq 1$ while $e\varphi/\left(e\varphi+mc^2\right)<1$. Therefore, as soon as we choose a positive energy solution $E=+\sqrt{m^2c^4+p^2c^2}$ for a freely moving particle far from the source of electric field, the probability density for the particle remains positive in the entire space, including the region near point-like source.

By such a way we have shown that the re-definition of the energy and momentum operators (7a-b) instead of corresponding customary definitions (1a-b) ensures physically meaningful solutions regarding the known problem of a moving charged particle in a vicinity of immovable source of electric field. Historically, this problem occurred crucial for rejection from the Klein-Gordon equation in the description of charged scalar particles and limiting its application to scalar fields only [4]. Thus, the meaningful solution (35) of this problem makes topical further re-analysis of the Klein-Gordon equation and its consequences in application to the description of scalar particles.

## 4. Conclusion

In section 2 we have shown how the analysis of quantum phase effects for moving dipoles and point-like charges [11-15, 22] based on the superposition principle for quantum phases allowed us to disclose two new quantum phases for point-like charges (12), (13), named the complementary electric and complementary magnetic A-B phases, correspondingly [11, 12].

Next, we have shown that the inclusion of complementary phases (12), (13) into the solutions of the fundamental equations of quantum mechanics requires to abandon the customary definition of the momentum operator (5) and to re-define this operator in the form (6). In order to preserve the gauge-invariance and Lorentz invariance in the relativistic case, the re-definition of the momentum operator should be done conjointly with the corresponding re-definition of the energy operator, see eqs. (7a-c).

We emphasized that the experimental discoveries of the A-B phase [17] for point-like charges and A-C and HMW phases [6, 7] for dipoles unambiguously prove the validity of the SQP principle and do validate subsequent new definitions of energy and momentum operators (7a-b), that makes topical the re-analysis of the fundamental equations of quantum mechanics with operators (7a-b).

In section 3, we applied the new energy and momentum operators (7a-b) to the Klein-Gordon equation with respect to a particular – but historically sound – problem, dealing with the motion of a spinless charged particle in a vicinity of a stationary point-like source of an electric field. As is known, the solution of this problem with the application of standard energy and momentum operators (1a-b) is given by eqs. (3, 4); the latter, in general, is meaningless from the physical viewpoint, since it admits solutions with a negative probability density for a moving charge. As is known, such a result evoked the necessity of giving up the application of the Klein-Gordon equation to scalar charged particles, and to limit the area of its applicability to scalar fields only [4].

Now, we have shown that the application of the new energy and momentum operators (7a-b) allows modifying the solution of the Klein-Gordon equation to the form (35), which guarantees a positive probability density of charged particle in the entire space. This result substantiates the applicability of this equation to scalar charged particles, and makes topical re-analysis of its other known difficulties, basically related to the fact that the Klein-Gordon equation contains the second order time derivative of the wave function [3, 23].

Such a re-analysis lies beyond the scope of the present paper, though we hope that the results obtained above may stimulate its performance from a new perspective related to the re-definition of the energy and momentum operators (7a-b).

## Authors' contributions

## Competing interests

## Funding
We received no funding for this study.